\documentclass[fleqn]{llncs}
\usepackage{times}
\usepackage{url}
\usepackage{xcolor}
\usepackage{soul}
\usepackage[utf8]{inputenc}
\usepackage{amssymb}
\usepackage{amsmath}
\usepackage{amsfonts}
\usepackage{graphicx}

\newcommand{\as}{``}

\newcommand{\be}{\begin{em}}
	\newcommand{\ee}{\end{em}}
\newcommand{\bb}{\begin{bf}}
	\newcommand{\eb}{\end{bf}}

\newcommand{\tbs}{\hspace*{4mm}}

\newcommand{\tbl}{\hspace*{12mm}}

\newcommand{\meno}{\medskip\noindent}

\newcommand{\rif}{~\ref}
\newcommand{\IF}{\mbox{:-}}
\newcommand{\op}{\mathit{Op}}
\newcommand{\THEN}{\mbox{${:}{>}$}}
\newcommand{\WH}{\mbox{${:}{<}$}}

\newcommand{\G}{G}
\newcommand{\F}{F}

\newcommand{\N}{N}

\newcommand{\ii}{{\cal{I}}}
\newcommand{\la}{{\cal{L}}}

\newcommand{\ag}{{\mathit{Ag}}}

\newcommand{\maag}{{\cal{M}}^{\ag}}

\newcommand{\repair}{\eta}

\newcommand{\react}{\rho}

\newcommand{\calsevp}{{{\cal{S}}^{{\cal{E}}vp}}}

\newcommand{\calsf}{{{\cal{S}}^{{\cal{F}}}}}

\newcommand{\calsj}{{{\cal{J}}^{{\cal{J}}}}}

\newcommand{\PRE}{\mbox{\,$:<$\,}}

\newcommand{\s}{\mathit{solve}}
\newcommand{\sn}{\mathit{solve\_not}}
\newcommand{\prol}{:\!\!-} 
\newcommand{\upp}{\mbox{$\up$}}

\newcommand{\up}{\uparrow}

\title{Ensuring Trustworthy and Ethical Behaviour\\in Intelligent Logical Agents\thanks{Copyright \copyright 2020 for this paper by its authors. Use permitted under Creative Commons License Attribution 4.0 International (CC BY 4.0).}}
\vspace{-0.5cm}
\author{
	\ \ Stefania Costantini
}

\institute{
	\begin{tabular}{lc}
		& Universit{\`{a}} degli Studi di L'Aquila \\
		& Dipartimento di Ingegneria e Scienze dell'Informazione, e Matematica\\
		& Via Vetoio snc, Loc. Coppito, I-67010 L'Aquila - Italy \\
		& {\tt stefania.costantini@univaq.it}
	\end{tabular}
}

\begin{document}
	\begin{sloppypar}
		
\maketitle
		

\vspace{-0.5cm}
\begin{abstract}
  Autonomous Intelligent Agents are employed in many applications upon which the life and welfare of living beings and vital social functions may depend. Therefore, agents should be \emph{trustworthy}. A priori certification techniques (i.e., techniques applied prior to system's deployment) can be useful, but are not sufficient for agents that evolve, and thus modify their epistemic and belief state, and for open Multi-Agent Systems, where heterogeneous agents can join or leave the system at any stage of its operation. In this paper, we propose/refine/extend dynamic (runtime) logic-based self-checking techniques, devised in order to be able to ensure agents' trustworthy and ethical behaviour.
\end{abstract}

\section{Introduction}

 The development, refinement, implementation and integration of methods for implementing Intelligent Agents so as to ensure transparent, explainable, reliable and ethical behaviour is strongly needed. This is due to the fact that agent systems are being widely adopted for many important autonomous applications upon which the life and welfare of living beings and vital social functions may depend. Therefore, agents should be trustworthy in the sense that they could be relied upon to do what is expected of them, while not exhibiting unwanted behaviour. So, agents \emph{should not} behave in improper/forbidden/unethical ways given the present context, and they \emph{should not} devise new behaviours that might be in contrast with their specification or however with the user's expectations. They should be transparent, in the sense of being able to explain their actions and choices when required: in fact, it should always be possible to find out how and why an agent (or, more generally, an autonomous system) made a particular decision. This property is not guaranteed by default, rather it descends from careful design methodologies. Transparency (\as explainability'') is vital since, in case of any kind of accident or misfunctioning involving an autonomous system, it must always be possible that the faults or inadequacies that caused the problem be identified and fixed, and accountability established. Moreover, understandably, users would (rationally) trust more those autonomous systems that could provide an intelligible explanation of their behaviours and choices. Numerous studies have linked users’ trust to the degree of confidence to be interacting with a system that is verifiable and can provide explanations which are intelligible, to each specific category of users. 
 
 Agents should report to their users in case the interaction with the environment results in the identification of new objectives to pursue. In fact, as cleverly observed by Stuart Russel in his recent book \cite{Russel2019}, the continuous interaction of an agent with the external environment may lead the agent to acquire new knowledge and also, based on this knowledge, to develop new objectives (or new plans for existing objectives) that might not be in line with the ethical principles that the human designer believed to have instilled into the agent, and might even violate basic human principles and rights. To avoid this, agents' behaviour should be \emph{verified} in a rigorous way, possibly integrating different verification methods, with the aim to ensure adherence of agents' operation to their specification. Such verification should, at least partly, happen at run-time, in order to monitor the agent's dynamic behaviours.
 
 In this paper, we discuss the issue of agents' verification, and its application to ethics, i.e., to making agent systems trustworthy w.r.t. their expected ethical behaviour. We propose some technical contributions concerning run-time verification, restricting ourselves to agent systems based upon computational logic. Many computational-logic-based agent-oriented languages and frameworks to specify agents and Multi-Agent Systems (MAS) have indeed been defined over time (for a survey of these languages and architectures the reader may refer, among many, to \cite{BordiniBDFGLOPR06,GarroMTBBBT19,CalegariCMO21}). Their added value with respect to non-logical approaches is to provide clean semantics, readability and verifiability, as well as transparency and explainability `by design' (or almost), as logical proofs can easily be transposed into natural-language explanations. All these approaches  are based (more or less directly, more or less adherently) on the so-called BDI ('Belief, Desires, Intention') model of agency \cite{RG91}, which formalizes means-end-reasoning and thus encompasses the concepts of perception, goals, actions, plans, commitments. The BDI model is inspired by Bratman's theory of human practical reasoning \cite{Bratman2009}, so such concepts are considered \emph{mental attitudes}, or \emph{mental/epistemic states} of agents.

Pre-deployment verification methods for logical agents (also called `static', or `a priori') are able to certify that agents will fulfil certain requisites of trustworthiness; this means that they will \emph{do} what is expected from them, and they will \emph{not violate} certain rules of behaviour. However, this kind of verification can be not fully sufficient for agents that will revise their beliefs and objectives in consequence of the interaction with a changing not-always-predictable environment. Dynamic, or Runtime, Verification (RV) is the \as orthogonal'' approach, aimed to verify whether a software component under observation respects or not, over time, during its operation (i.e., at `runtime'), some given properties. Approaches to dynamic verification are meant to be \emph{lightweight}, so as not to be a burden on the system's performances. The two kinds of verification methods can be profitably exploited in conjunction, as both approaches do not examine all possible system’s behaviours, but only some explicitly designated ones, as specified by the system's designer.

The work presented in the present paper stems from the observation that, in many practical cases, in particular in agents that learn, or in open MAS where agents can join and leave the system, it is not possible to fully predict the set of events that will be perceived and considered by an agent, which might even lead the agent itself to devise new objectives and plans that depart from its expected/acceptable behaviour. 
We advocate that {\bf agents themselves should keep their own operation under control in changing circumstances}: i.e., agents should be able {\bf to observe and to check their own behaviour}, {\bf to take countermeasures against anomalies} and also, very importantly, {\bf to correct and constantly improve such behaviour}, so as, for instance, to refine their understanding of human preferences and expectations, as required in \cite{Russel2019}. Without that, agents' behaviour may become dangerously unpredictable. In our view, agents should thus {\bf reflect} on themselves, and act accordingly to modify their state and/or their interaction with the humans and the environment. 
The methods that we propose fall within the RV techniques, although our notion of self-observation and self-improvement is new in this field.

We find similarities between our approach and the point of view of \emph{Self-aware computing}: quoting \cite{TorrePY15},
\emph{Self-aware and self-expressive computing describes an emerging paradigm for systems and applications that proactively gather information; maintain knowledge about their own internal states and environments; and then use this knowledge to reason about behaviours, revise self-imposed goals, and self-adapt. $\ldots$ Systems that gather unpredictable input data while responding and self-adapting in uncertain environments are transforming our relationship with and use of computers.} Reporting from \cite{Amir2007}, \emph{From an autonomous agent view, a self-aware system must have sensors, effectors, memory (including representation of state), conflict detection and handling,
reasoning, learning, goal setting, and explicit awareness
of any assumptions. The system should be reactive,
deliberative, and {\bf reflective}.} 

An example of application of these concepts, devised for computational-logic-based agents, is presented in \cite{PerlisJLC2005}, which defines a time-based \emph{active logic}
and a \emph{Metacognitive Loop} (MCL): such loop specifies the system's self-monitoring, via
reasoning and meta-reasoning, with self-correction whenever needed. As discussed in \cite{PerlisJLC2005},
MCL continuously monitors an agent's expectations,
notices when they are violated,
assesses the cause of the violation
and guides the system to an appropriate response. In the terms of \cite{Amir2007},
this is an example of \emph{Explicit Self-Awareness}, where the computer system has a full-fledged self-model representing knowledge about itself.

In this paper, we propose contributions to an envisaged toolkit for run-time self-monitoring of evolving agents. Differently from \cite{PerlisJLC2005} however, we do not aim to continuously monitor the entire system's state, but rather to monitor only the aspects that a designer deems to be relevant for keeping the system's behaviour safely under control. We specify techniques and tools for: (i) checking the immediate, \as instinctive'' reactive behaviour in a context-dependent way, and (ii) checking and re-organizing an agent's operation at a more global level.
In particular, we introduce meta-rules and meta-constraints for agents' run-time self-checking, where checking of each specified condition occurs upon need, or at a certain --customizable-- frequency.
The proposed meta-constraints are based upon a simple interval temporal logic tailored to the agent realm, which we called
A-ILTL (`Agent-Oriented Interval LTL')\footnote{LTL stands as customary for `Linear Temporal Logic'. For an introduction and formal definitions concerning Temporal Logics cf., e.g., \cite{tl3}.}. A-ILTL constraints and evolutionary expressions are defined over formulas of an underlying logic language
${\cal L}$, where however we make A-ILTL independent of ${\cal L}$, thus ensuring applicability of our approach to any logic-based language.

In A-ILTL, a designer can specify properties that should hold in specific time instants and time intervals, according to past but also future events.
\emph{Evolutionary A-ILTL Expressions}, that we have implemented and experimented, are in fact composed of
the following elements. (i) A (partially specified, possibly empty) sequence of events that have happened (i.e., have been perceived by the agent); the occurrence of an instance of such sequence enables the check of (ii) a temporal-logic-like
expression defining a property that should hold (in a given interval), provided that the agent monitors (iii) a (possibly empty) sequence of events that are supposed to happen in the future, without affecting the property, or that are supposed not to happen, otherwise the property is no longer significant; and, finally,
(iv) \as repair''/\as improvement'' countermeasures (optional) to be undertaken if the property is violated. 
Countermeasures can be: at the object-level, i.e., related to the application domain at hand; or, at the meta-level, e.g., they can inspect elements of the agent's internal state, and even result in replacing a software component with a diverse alternative. The act of checking A-ILTL expressions can indeed be considered as an introspective act, as an agent suspends its current activities in order to inspect and possibly self-modify its own state. 

 Our work has a clear connection in its objectives to the work of \cite{GiacomoIFP19}, which proposes to implement a \as restraining bolt''\footnote{A \as restraining bolt'' as imagined in the Star Wars Science Fiction saga is a small cylindrical device that, when activated, restricts a droid's actions to a set of behaviours considered desirable/acceptable by its owners.} for agents' behaviour; this by conditioning reinforcement learning of reactive actions, so as to obey LTL specifications defining behavioural/ethical principles. This (very promising) method is orthogonal to ours, thus the two might profitably coexist.
 
A toolkit for logical agents' run-time self-verification can be obtained by means of the synergy between the new features proposed here and those introduced in past work (notably \cite{rcra2010,LANMR2012}, \cite{AAMAS2013}, and \cite{CostantiniGDP18y})\footnote{The author acknowledges the co-authors of the aforementioned papers, that contributed to various extents to the development of this research. Apart from \cite{AAMAS2013} which is an extended abstract, all these other papers appeared in venues with a limited audience and/or without formal proceedings.}.
The proposed approach can be seen under two perspectives. On the one hand, as a means of defining an enhanced \as restraining bolt'' (not in exclusion but complementary to the one of \cite{GiacomoIFP19}), capable of preventing agents from engaging in unwanted behaviours which could not be fully predicted at design time. On the other hand, since agents are supposed to be able to learn rules of behaviour over time, as a means of defining a potentially \as disobeying robot'' that can on occasion disallow behaviour hardwired at design time; this in cases where, in the present agent's context, such behaviour violates context-dependent learned behavioural or ethical rules\footnote{There is an open discussion in the literature, initiated by Arkin in \cite{Arkin2018}, about whether agents should be allowed not only to disobey, but also, on occasions, to deceive (though, according to Kant’s categorical imperative, lying is fundamentally wrong). This for cases where deception may have societal value, to preserve the user's state of mind or even possibilities of survival. A problem is, however, how to ensure that deception is only used in the contexts it was designed for. To this problem, the proposed approach might provide some initial answer.}.

The paper is organized as follows: in Sections\rif{relwork} and\rif{relworkethics} we introduce and discuss existing approaches to the verification of agent systems, and very shortly provide a review and some pointers to existing work on Machine Ethics, in particular concerning Logical Agents; in Section\rif{rp} we introduce metarules for checking reactive behaviour, and in Sections\rif{runtimecheck}, \rif{lsproperties} and\rif{beh-rul} we introduce A-ILTL constraints in theory and practice. In Section\rif{experiments} and \rif{complexity} we present some experiments, and discuss the complexity of our approach in terms of the burden that it might add to agents' execution performance. In Sections\rif{conclusions} we briefly discuss some more related work, and then conclude. In the examples, we adopt a Prolog-like syntax (cf. \cite{lloyd87}) for rules, of the form $\mathit{Head} \prol \mathit{Body}$, where $\mathit{Head}$ is an atom, $\mathit{Body}$ is a conjunction of literals (atoms or negated atoms, that are often called 'subgoals') and the comma stands for $\wedge$. The specific syntax that we use is however purely aimed at illustration: the approach can be, `mutatis mutandis', re-worked w.r.t. any different syntax.

\section{Background: Verification Methods for Agent Systems}
\label{relwork}

The verification of a MAS
global behaviour, as well as the verification of a single agent or, more generally, of an autonomous system, is an intrinsically complex but unavoidable problem: in fact, many different approaches to its solution have been presented in the literature. For an extensive illustration and comparison of approaches to ensuring reliability of autonomous systems, we refer the reader to the recent survey \cite{FisherMRSWY21} and to the many references therein.

In this section, we intend to shortly introduce, for the sake of completeness, the most established methods for the verification of agents or MAS, i.e.: (i) verification that is
performed statically, or `a priori' (prior to deployment) by checking the system against given input configurations, sequences of external events, etc.,
and (ii) verification that is performed dynamically, by monitoring the evolution of the system during its operation (at `runtime'), in order to stop, or try to recover, every situation that appears incorrect with respect to the system's specification. The methods that we propose in this paper are of the latter kind. 

Static verification can be accomplished through model checking \cite{Clarke:mc,EC2018}, abstract interpretation
(not commented here, cf. \cite{abstr-int}) or theorem proving \cite{leveque2010,Shap}. Although model checking techniques have been originally adopted for testing hardware devices, their application to software systems and protocols is constantly growing \cite{spin,mchMAS}, and there have been a number of attempts to overcome some known limitations of this approach, for instance so-called `state explosion', which occurs when the situations to check generate too many combinations. 

The application of static techniques such as model checking to the verification of MAS encounters some difficulties, due to the marked differences between the languages used for the definition of agents and those needed by verifiers (usually ad-hoc, tool-specific languages). The model-checking paradigm, in fact, allows one to model a system $S$ in terms of an automaton, by building an implementation $P_s$ of the system at hand by means of a model-checker-friendly language. Programs written in such suitable input language can
then be submitted to model checkers in order to verify formal
specifications. These are commonly expressed either as formulae of
the Branching Time Computational Tree Temporal Logic CTL \cite{modch,modch2} or as
formulae of Linear Temporal Logic LTL \cite{Holtz,Vardi,VardiR10,Rozier11}. 
It can be not easy to re-model an agent or MAS in another language: this task is usually performed (at least partly) manually, and thus it requires an advanced expertise and gives no guarantee on the correctness and coherence of the new model: the current research in this field is in part still focused on the problem of defining a suitable language that can be used to easily and/or automatically reformulate a MAS in order to verify it through general model checking algorithms (cf., e.g., \cite{mchMAS,walton}). The literature reports however fully-implemented verification frameworks (e.g., \cite{visser2,Lom2,sciff-iclp08,LomuscioQR17}). 

Lomuscio et al. have defined the
bounded semantics of CTLK \cite{Lom1,Lom2}, a combined logic of knowledge
and time. Their approach is to translate the system model and a
formula $\phi$, indicating the property to be verified, into sets of
propositional formulae to be then submitted to a SAT-solver. The approach has evolved over time until \cite{Lomuscio17,LomuscioQR17}, leading to the proposal of an open-source model checker, MCMAS, for the verification of MAS. MCMAS takes ISPL (Interpreted Systems Programming Language) descriptions as input, where an ISPL file fully describes a multi-agent system, i.e., both the agents and their environment. 
Model-checking
techniques have been adopted in order to check systems implemented
in AgentSpeak(L) \cite{FisherAg}, where a variation of the language
aimed at allowing its algorithmic verification has been proposed. The work \cite{visser2} proposes in fact a technique to model-check agents defined in a subset of the AgentSpeak language, that can be automatically translated into languages accepted by the model checkers SPIN \cite{spin} and Java PathFinder \cite{visser1}. These simplifications and translations - though partly or mostly automated - make the process of model checking, though very useful, not-too-easy to apply. 

Finally, concerning model checking of agent systems we mention the recent work presented in \cite{Dennis16,Dennis18,Dennis21} concerning the {MCAPL} model checking framework, where \cite{Dennis16} explicitly considers checking agents' ethical choices.

Techniques to reduce the heaviness of model-checking have been devised or exploited in the above-mentioned works, still, the amount of computational resources needed by model checking is considerable. 

The deductive approach to verification uses a logical formula to
describe all possible executions of the agent system that one wants to check, and then
attempts to perform theorem proving of a required property from this logical formula.
Such properties are often captured using modal and
temporal logics. Deductive approaches have been adopted by Shapiro,
Lesperance and Levesque that defined CASLve \cite{Shap}, a
verification environment for the Cognitive Agent Specification
Language. A limitation of the theorem proving approach is the problem's
complexity, and thus a human interaction is often
required.
In this field, the author of this paper has proposed (with others), in \cite{clima-XI-2010}, an approach
to the formal description of the operational semantics of any agent-oriented logic
language and of its underlying inference engine. We have fully formalized the DALI agent-oriented programming language \cite{jelia02,jelia04,postdalt06,daliDOI}
and its interpreter \cite{Toc05}. We have been able to prove various properties of the language
(e.g., properties of the communication protocols that DALI provides) and of its interpreter,
first of all correctness of the interpreter w.r.t. the procedural (resolution-based)
semantics of DALI, that can be used as basis to prove properties of DALI agents.

Overall, the question that a priori verification tries to answer is: \as given a set of rules that the agent will respect, are these rules enough to guarantee the desired future behaviour, independently of what will happen in an open environment?''
However, if an agent is supposed to learn new knowledge or rules, then
there can be properties that it is difficult or even impossible to fully check by means of the above techniques either a priori, or by repeating the check whenever the agent performs some learning. Moreover, a MAS can be composed of heterogeneous agents and it can be open, i.e., agents can freely join or leave the system. In this cases, a priori verification is clearly insufficient.

Another possible approach to agent validation, based on \emph{testing}, requires observing the agent's behaviour as it performs its tasks in a series of
test scenarios before putting it at work. This approach however, as
observed by Wallace in \cite{wallace}, is by its very nature
incomplete since all critical scenarios can hardly be identified and examined. 

So, it has been found useful to individuate mechanisms to complement a priori verification and testing, capable of verifying an agent's behaviour correctness without stopping its operation.  Dynamic, or runtime (RV) verification is the only (semi-)formal verification technique that directly analyzes system's operation to check for violations of formally expressed specifications/properties. This although, as remarked in \cite{Rozier16}, \as Specification of the requirements to monitor at runtime is the biggest bottleneck to successful deployment of RV''. This is an issue to which this paper will try to provide at least a partial answer.

A relevant recent approach to RV, based upon computational logic and especially tailored to logical agents, is that of \emph{Trace Expressions} (cf. e.g., \cite{FerrandoDA0M18,FerrandoWCDM19}), which  
are in fact a specification formalism especially devised to this aim.
An event trace, in this approach, is a (possibly infinite) sequence, defined over a fixed universe of events. A trace expression, built out of suitable constructs, denotes a specific set of event traces. These specify in a formal way which are the events allowed in a certain state for a given agent. A Prolog implementation has been devised to allow a user to automatically build a trace-expression-driven monitor (by means of a user-friendly language, currently under design); this monitor will be able to observe events taking place in the environment, and check whether such events are indeed allowed in the current agent's state. A system can successfully terminate if the trace expression representing the current state can halt, i.e., it contains the empty trace. Experiments demonstrate that, in most cases, verification of trace expressions is linear in the length of the trace, whatever available modelling features have been exploited; thus, performances are guaranteed to be acceptable. 

Interestingly, trace expressions can be exploited for use in a model checker: in fact, an algorithm has been proposed to check LTL properties satisfiability on trace expressions \cite{Ferrando19}. Vice versa there are tentative approaches, not yet applicable to agents but nonetheless interesting, notably \cite{MC-RV2018}, to adapting existing software model checkers to perform runtime verification. 

The difference with our approach is that we do not check event traces `per se'. Rather, we define constraints based upon a special interval logic, to specify which behaviours are allowed or not, also depending upon the present agent's BDI state: which is to say, in the formulation of these constraints it is possible to access meta-level notions about the agent's internal state such as goals, plans, actions; that is why our technique is `introspective'. Moreover, our constraints can (optionally) take a sequence of events that are supposed to have happened as a precondition, i.e., a certain behaviour is allowed or required after certain events sequences; but, we also consider events which are expected or prohibited in the future, to evaluate whether a constraint has succeeded (is satisfied) or not. Another main difference is that we do not devise a monitor which is conceptually external to the agent: rather, the proposed meta-level rules and constraints act as monitors, although fully integrated into the agent's operation.

\section{Background: Machine Ethics, Ethics in Agents and MAS}
\label{relworkethics}

AI ethics, or – more generally – Machine Ethics, is concerned with the question of how AI systems can behave ethically. It is a recent research field, dating back to the early 2000's, concerning both philosophy and computer science. In fact, Philosophers should answer questions on whether a machine could behave ethically, and on the basis of which ethical principles, and are challenged with the more general question on whether society should (and to which extent) delegate moral responsibility to machines. Computer scientists are concerned with devising techniques usable to build ethical machines. In the definition of such techniques, the distinction must be made between implicitly ethical agents, i.e., (i) machines designed to avoid unethical outcomes, and (ii) explicitly ethical agents, i.e., machines that can reason about ethics. In this paper, we propose an approach that copes with case (i), and only to some extent with case (ii).

According to Winfield \cite{WinfieldMPE19}, where an extensive discussion with many references can be found, the field of machine ethics was established by Allen et al. \cite{Allen2000,Allen2005}, where the concept of \as Artificial Moral Agent'' was introduced, and three approaches to machine ethics were identified: top-down, bottom-up and hybrid (combination of top-down and bottom-up). The top-down approach constrains the actions of the machine in accordance with pre-defined rules or norms. The bottom-up approach instead requires an agent to be able to learn, recognise and correctly respond to morally challenging situations. Other relevant contributions soon followed (cf. among many, \cite{Asaro2006,Moor2006,Powers2006,AndersonAA06}). The proposal by Arkin \cite{Arkin2018} for the design and implementation of an \emph{ethical governor} for robots -- intended as a run-time mechanism for moderating or inhibiting a robot’s behaviour to prevent it from acting unethically – brings a conceptual similarity with the approach proposed in the present paper. Because of the burden of expectation on the behaviour of ethical machines, such \emph{governor} will need to be especially robust, and to this aspect our work attempts to provide some contributions.

To date, many approaches exist to Machine Ethics, Ethics in Artificial Intelligence systems, and more specifically Ethics in Agents and MAS, where the latter ones are of particular concern for the topics treated in this paper. For recent literature reviews, a reader may refer to  \cite{DyoubThesis2019,Nallur2020,CosDL2020,TolmeijerKSCB21}. There, the authors start from the illustration of moral philosophical concepts ranging from ancient philosophers to recent works in neurology
and cognitive sciences, discuss concepts like morals, ethics,
judgment or values, and then identify the kinds of philosophical ethical theories that have been applied or are potentially applicable to agents and multi-agent systems, providing many relevant references. In particular, \cite{DyoubThesis2019,CosDL2020} concern theories developed in logic, and then transposed into Computational Logic. They discuss in some depth two seminal lines of work: the one by Pereira et al., starting from the famous book \cite{PereiraS16}, and summarized in the more recent book \cite{PereiraL20}, which exploits a blend of many forms of logic programming; and, the one by Marek Sergot (cf. \cite{Sergot06,Sergot07,Sergot09,Sergot10}), mainly exploiting Answer Set Programming to compare and weigh alternative scenarios in order make ethical decisions\footnote{Answer Set Programming (ASP) is a successfully logic 
	programming paradigm (cf. \cite{ASPJournal2016} and the references therein) stemming from the Answer Set (or \as Stable Model'') semantics of Gelfond and Lifschitz \cite{gelfond1988stable,gelfond1991classical}, and based on the programming methodology proposed by
	Marek, Truszczy{\'{n}}ski and Lifschitz \cite{marek1999stable,Lifschitz99}. ASP is put into practice by means of effective inference engines, called \emph{solvers}, which are freely available, see \cite{ijcai2018-769}.}. 

Other more recent attempts at modelling and implementing forms of Ethical Reasoning in logical agents are \cite{Cointe16,Berreby17,Berreby18} that try to exploit Answer Set Programming and (variants of) the Event Calculus \cite{event_calculus86}. A very recent work \cite{DennisB0021} develops theory and concepts concerning ethical reasoning in changing contexts.

Ethical rules to be exploited in ethical reasoning can be defined a priori by a system's designer, but, in alternative, they can be learnt, as proposed in \cite{CosDL2019,DyoubCL19,DyoubCL19b,DyoubCLL20}.

However, the above-mentioned approaches propose theories and implementations to represent and reason about ethical principles and their applications, i.e., how agents should make ethical judgements and decisions. The tools proposed in this paper aim to check and possibly enforce respect of such decisions, so they are \as agnostic'' w.r.t. the kind of ethical reasoning that is performed. That is why we do not go into any further depth concerning Machine Ethics.

\section{Checking Agents' Reactive Behaviour}
\label{rp}

In the BDI model, an agent will have objectives, and devise plans to reach these objectives.
In addition, most agent-oriented languages and frameworks provide mechanisms for `pure' reactivity, i.e. `instinctive' immediate reaction to an event. The acceptable reactions that an agent can enact are in general strictly dependent on the context, the agent's role and the situation. Let us consider the need to ensure an agent's ethical behaviour. Assume for sake of example an agent (either human or artificial) which finds itself to face some other agent which might be, or certainly is, a criminal. We can state in general that: (i) if the context is that of playing a video-game, every kind of reaction is allowed, including beating, shooting or killing the `enemy', with exceptions, e.g., when small children are watching; (ii) same if the context is a role game: the players can pretend to threaten, shout or kill the other players, where every action is simulated and thus harmless; (iii) in reality, a citizen can shout, call the police, and try enact defensive strategies or actions; a policewoman/policemen can threaten, arrest, or in extreme cases shoot the suspect criminal.

Or, assume that a self-driving car has to decide on whether to accelerate or not, where accelerating is in general allowed if the speed limit has not been reached. The different contexts here may concern whether the car is in town, or out of town, or on a motorway, as in each of these cases the speed limit is different. There are however exceptions, due to speed limitations that can be found on the way for various reasons (e.g., construction), or due to the kind of vehicle, as for instance an ambulance, or the police, or the fire truck, can run faster than the limit in case of an emergency.

The reaction to enact in each situation can be `hardwired' by the agent's designer. In the case of the self-driving car, it might seem that a priori verification could be sufficient; however, if one considers the unpredictability of circumstances (i.e., when and where speed limits can be found or emergencies can arise), a blend with dynamic verification can be more practical: this case is in fact discussed in \cite{FerrandoDA0M18}, and coped with by means of trace expressions.

In the other case, general ethical rules will reasonably be provided, where however their specific application in the domain at hand can be learned, e.g., via reinforcement learning. 
Here, run-time checking of agent's behaviour w.r.t. ethical rules is in order, as the results of learning are in general unpredictable and to some extent potentially unreliable. The method of conditioning reinforcement learning to obey desirable properties proposed in \cite{GiacomoIFP19} is applicable but might not completely suffice, due to possible unexpected dynamic changes of context and roles not foreseeable in advance.

In this section, we introduce mechanisms to verify and possibly enforce desired properties of reactive behaviour by means of metalevel rules. To define such new rules, we assume to augment the underlying language ${\cal L}$ at hand with a naming (or \as reification'') mechanism, and with the introduction of two
distinguished predicates, $\s$ and $\sn$. These are meta-predicates, that can be employed to control the object-level behaviour.
In fact, $\s$, applied to (the name of) an atom which represents an action or an objective of an agent, may specify conditions for that action/goal to be enacted;  vice-versa $\sn$ specifies under which conditions it should be blocked.

Below is a simple example of the use of $\s$: the aim is to specify that action $\mathit{Act}$ can be executed in the present agent's context of operation $C$ and role $R$, only if this action is allowed, and it is deemed to be ethical w.r.t. context and role. Any kind of reasoning can be performed via metalevel rules in order to monitor and assess base-level ethical behaviour. Below, lowercase syntactic elements such as $p'$, $c'$, are \emph{names} of predicates and constants, according to some \emph{naming mechanism}\footnote{A \as naming relation'', or \as naming mechanism'' or \as {reification mechanism}''
	is a method for representing, within a first-order language, expressions of the language itself
	without resorting to higher-order features.
	Naming relations can be introduced in several manners.
	For a discussion of different possibilities, with their advantages
	and disadvantages, see, e.g., \cite{hill+lloyd:meta88,barklund:meta88,harmelen:meta92,bcdl:propEnc}.
	However, all such mechanisms are based upon introducing distinguished constants, function symbols (if available), and predicates,
	devised to construct names.
	For example, given an atom $p(a,b,c)$, a name might be $\mathit{atom(pred(p'),args([a',b',c']))}$
	where $p'$ and $a',b',c'$ are new constants intended as names for the syntactic elements $p$ and $a,b,c$.
	Notice that: where $p$ is a predicate symbol, which is not a first-class object in a first-order setting, its name $p'$ is a constant, which is instead a first-class object and can be manipulated. The possibility to manipulate, even if indirectly, every syntactic object of given language is the purpose of the introduction of names. In the above sample name,
	$\mathit{atom}$ is a distinguished predicate symbol, $\mathit{args}$ a distinguished function symbol and $[\ldots]$ is a list. This name might be shortened as $p'(a',b',c')$. Naming mechanisms have been widely studied, cf., among many, \cite{PerlisS94,BarklundDCL95-1,CostantiniMetarev02}. Whatever the chosen naming mechanism, it is necessary to relate objects and their names. It is common practice to denote the name of an object $\alpha$ as $\up \alpha$. E.g., $\up p(a,b,c)$ = $\up p(\up a, \up b, \up c)$. Since in our sample naming relation we have stated that $\up p = p'$, $\up a = a'$, $\up b = b'$, and $\up c = c'$, we have $\up p(a,b,c) = p'(a',b',c')$.}, and uppercase syntactic elements such as $V'$ are metavariables. 

$
\begin{array}{l}
\s(\mathit{execute\_action}'(\mathit{Act}')) \prol\\ 
\tbl\mathit{present\_context(C)},\mathit{agent\_role(R)},\\
\tbl\mathit{allowed(C,R,Act')},\mathit{ethical(C,R,Act')}.
\end{array}
$

\smallskip
 We assume that $\s(\mathit{execute\_action}'(\mathit{Act}'))$ is automatically invoked whenever subgoal (atom) $\mathit{execute\_action}(\mathit{Act})$ is attempted at the object level. More generally, given any subgoal $A$ at the object level, if there exists an applicable $\s$ rule, then such rule is automatically applied, and $A$ can succeed only if $\s(\upp A)$ succeeds, where the expression $\upp A$ denotes the \emph{name} of $A$ according to the chosen naming mechanism. We assume, also, that the present context and the agent's role are kept in the agent's knowledge base. Since both parameters may change, the same action may be allowed in some circumstances and not in others. Notice that the predicate $\mathit{ethical}$ is meant to be user-defined because, as said before, our approach is agnostic w.r.t. the ethical principles that an agent's designer intends to enact. 
 
Symmetrically we can define metarules to forbid unwanted object-level behaviour, e.g.:

$
\begin{array}{l}
\sn(\mathit{execute\_action}'(\mathit{Act}')) \prol\\ \tbl\mathit{\mathit{present\_context(C)},\mathit{ethical\_exception(C,Act')}}.
\end{array}
$

\noindent this rule prevents successful execution of its argument, in the example $\mathit{execute\_action}(\mathit{Act})$, whenever $\sn(\upp A)$ succeeds. Then, action/goal $A$ can succeed (according to its object-level definition) only if $\s(\upp A)$ (if defined) succeeds and $\sn(\upp A)$ (if defined) does not succeed.

The outlined functioning corresponds to \emph{upward reflection} when the current subgoal $A$ is \emph{reified} (i.e., its name is computed) and applicable $\s$ and $\sn$ metarules are searched; if such metarules are found, control in fact shifts from the object to the metalevel (consider that $\s$ and $\sn$ can rely upon any set of auxiliary metalevel rules). If no rule is found or whenever $\s$ and $\sn$ metarules complete their execution, \emph{downward reflection} returns control to the object level, to execution of $A$ if confirmed or to subsequent goals/actions if $A$ has been cancelled by either failure of an applicable $\s$ metarule or success of an applicable $\sn$ metarule.

Via $\s$ and $\sn$ metarules, fine-grained activities of an agent can be punctually checked and thus allowed and disallowed, according to the
context an agent is presently involved into with a certain role. 

Semantics of the proposed approach can be sketched as follows (a full semantic definition can be found in \cite{CostantiniL89,CostantiniL90,CostantiniF19}).
According to \cite{Dix95a}, in general terms we understand a semantics $\mathit{SEM}$ for logic knowledge representation languages/formalisms as a function
which associates a theory/program with a set of sets of atoms, which constitute the intended meaning. When saying that $P$ is a program, we mean that it is a program/theory in the (here unspecified) logic languages/formalism that one wishes to adopt.

We introduce the following restriction on sets of atoms $I$ that should be considered for the application of $\mathit{SEM}$. First, as customary we only consider sets of atoms $I$ composed of atoms occurring in the `ground' version of $P$ (where the ground version of program $P$ is obtained by substituting in all possible ways variables occurring in $P$ by constants also occurring in $P$). In our case, metavariables occurring in an atom must be substituted by metaconstants, with the following obvious restrictions: a metavariable occurring in the predicate position must be substituted by a metaconstant representing a predicate; a metavariable occurring in the function position must be substituted by a metaconstant representing a function; a metavariable occurring in the position corresponding to a constant must be substituted by a metaconstant representing a constant. According to well-established terminology, we therefore require $I \subseteq B_P$, where $B_P$ is the \emph{Herbrand Base} of $P$.
Then, we pose some more substantial requirements: we restrict $\mathit{SEM}$ to determine only \emph{acceptable} sets of atoms\footnote{modulo bijection: i.e., $\mathit{SEM}$ can be allowed to produce sets of atoms which are in one-to-one correspondence with acceptable sets of atoms},
where $I$ is an \emph{acceptable} set of atoms iff $I$ satisfies the axiom schemata:\\ 
\tbl\tbl\tbs\(A \leftarrow \s(\upp A)\ \ \ \ \neg A \leftarrow \sn(\upp A)\)

So, by means of this restriction we model the implementation of properties that have been defined via $\s$ and $\sn$ rules, without modifications to $\mathit{SEM}$. For clarity however, one can assume to filter away $\s$ and $\sn$ atoms from acceptable sets. In fact, the \emph{Base version} $I^B$ of an acceptable set $I$ can be obtained by omitting from $I$ all atoms of the form $\s(\upp A)$ and $\sn(\upp A)$.
Procedural semantics, and the specific naming relation that one intends to use, remain to be defined, where the 
above-introduced semantic modelling is independent of 
these aspects. For approaches based upon (variants of) Resolution (like, e.g., Prolog and like many agent-oriented languages such as, e.g., AgentSpeak \cite{Rao96}, {GOAL} \cite{GOAL2009}, 3{APL} \cite{DastaniRM05} and {DALI} \cite{jelia02,jelia04,postdalt06,daliDOI}) one can extend their proof procedure so as to automatically invoke metarules whenever applicable, to validate or invalidate success of subgoal $A$. 

How to define the predicate $\mathit{ethical(C,R,Act')}$? Again, rules defining this predicate can be specified at design time, or they can be learned, or a combination of both options. In previous works \cite{ADSCFL2019,ADILP2019,ADICLP2019}, a hybrid logic-based approach was proposed for ethical evaluation of agents' behaviour, with reference to dialogue agents (so-called `chatbots') but easily extendable to other kinds of agents and of applications. The approach is based on logic programming as a knowledge representation and reasoning language, and on Inductive Logic Programming (ILP) for learning rules needed for ethical evaluation and reasoning, taking as a starting point general ethical guidelines related to a context; the learning phase starts from a set of annotated cases, but the system is then able to perform continuous incremental learning.

\section{A Logic for Checking Agent's Behaviour over Time}
\label{runtimecheck}
The techniques illustrated in the previous section are \as punctual'', in the sense that they
provide context-based mechanisms to allow/disallow agents' actions. However, it is necessary 
to introduce ways to monitor an agent's behaviour in a more extensive way. 
In fact, properties that one wants to verify often depend upon time and time intervals, and possibly on
which events have been observed by an agent up to a certain point, and which others
are supposed to occur later. The definition of frameworks such as the one that we propose here,
for checking agent's operation during its `life' based on its experience and expectations, has not been widely treated so far in the literature.

Below we introduce a logic which constitutes the basis of our approach for checking an agent's behaviour
during the agent's activity. 

\subsection{A-ILTL}
\label{ailtl}

For defining properties that are supposed to be respected
by an evolving system, a well-established approach is that of
Temporal Logic, and in particular of Linear-time
Temporal Logic (LTL). This logic \cite{tl3}
evaluates each formula with respect to a vertex-labeled infinite path (or \as state sequence'')
$s_{0}s_{1}\ldots$ where each vertex $s_{i}$
in the path corresponds to a point in time (or \as time instant'' or \as state'').
In what follows, we use the standard notation
for the best-known LTL operators.

In \cite{LANMR2012}, we
formally introduced an extension to LTL
based on \emph{intervals}, called A-ILTL for `Agent-Oriented Interval LTL'.
A-ILTL is useful
because the underlying discrete linear model of time and the complexity of the logic remains unchanged with respect to LTL.
This simple formulation can be efficiently implemented, and is
sufficient for expressing and checking a number of interesting properties of agent systems.
Formal syntax and semantics of a number of
A-ILTL operators (also called below \as Interval Operators'') are fully defined in \cite{LANMR2012}.

LTL and A-ILTL expressions are interpreted in a discrete, linear
model of time. Formally, this structure is represented by ${\cal{M}} = \langle \mathbb{N}, \ii \rangle$
where, given countable set $\Sigma$ of atomic propositions, interpretation function
$\ii\,:\,\mathbb{N}$\,$\mapsto$\,${{2^{\Sigma}}}$ maps each natural number $i$ (representing state $s_i$) to a
subset of $\Sigma$. Given set $\cal{F}$ of formulas
built out of classical connectives and of temporal operators,
the semantics of a temporal formula is provided by the
satisfaction relation $\models\,:\,{\cal{M}}\,\times\,\mathbb{N}\,\times\,\cal{F}\,\rightarrow\,\{\mathit{true},\mathit{false}\}$.
For $\varphi \in \cal{F}$ and $i \in \mathbb{N}$ we write ${\cal{M}},i\,\models\,\varphi$ if,
in the satisfaction relation, $\varphi$ is true w.r.t. ${\cal{M}},i$.
We can also say (leaving ${\cal{M}}$ implicit) that $\varphi$ \emph{holds} at $i$, or equivalently in state $s_i$,
or that state $s_i$ satisfies $\varphi$.
For atomic proposition $p \in \Sigma$, we have ${\cal{M}},i$\,$\models$\,$p$ iff $p \in \ii(i)$.
The semantics of $\models$ for classical connectives is as expected, and the semantics for LTL operators is as reported in \cite{tl3}.
A structure ${\cal{M}} = \langle \mathbb{N}, \ii \rangle$ is a model of $\varphi$ if
${\cal{M}},i\,\models\,\varphi$ for some $i \in \mathbb{N}$.
Similarly to classical logic, a LTL or A-ILTL formula $\varphi$ can be
satisfiable, unsatisfiable or valid and one can define the notions of
entailment and equivalence between two formulas.

Some among the A-ILTL operators are the following, where $\varphi$ is an expression in an underlying agent-oriented language $\la$, and $m,n$ are positive integer numbers used to (optionally) specify the interval where the formula must hold, according to the semantics specified below. If the interval is not specified, then the meaning is the same as for LTL. A limitation that we impose is that temporal operators cannot be nested.

	\noindent
	$\F_{m,n}$ ({\em eventually (or \as finally'') in time interval}). $\F_{m,n}\varphi$ states that $\varphi$ has to hold sometime on the path from state $s_m$ to state $s_n$. I.e., ${\cal{M}},i\,\models\,\F_{m,n}\varphi$ if there exists $j$ such that $j \geq m $ and $j \leq n$ and
	${\cal{M}},j\,\models\,\varphi$.
	
	\noindent
	$\G_{m,n}$ ({\em always in time interval}). $\G_{m,n}\varphi$ states that $\varphi$ should become true at most at state $s_m$ and then hold at least until state $s_n$. I.e., ${\cal{M}},i\,\models\,\G_{m,n}\varphi$ if for all $j$ such that $j \geq m $ and $j \leq n$
	${\cal{M}},j\,\models\,\varphi$. Can be customized into $\G_{m}$, {\em bounded always}, where $\varphi$ should become true at most at state $s_m$.

	\noindent
	$\N_{m,n}$ ({\em never in time interval}). $\N_{m,n}\varphi$ states that $\varphi$ should not be true in any state between $s_m$ and $s_n$. I.e., ${\cal{M}},i\,\models\,\N_{m,n}\varphi$ if there not exists $j$ such that $j \geq m $ and $j \leq n$ and
	${\cal{M}},j\,\models\,\varphi$.
	
In practical use, as seen below A-ILTL operators will allow one to construct useful run-time constraints.

\subsection{A-ILTL and Evolutionary Semantics}
\label{a-iltl-evol}

In this section, we refine A-ILTL so as to operate on a sequence
of states that corresponds to the Evolutionary Semantics of an agent-oriented programming language \cite{CTpostdalt06}. This is a meta-semantic approach, as it is independent of the underlying agent-oriented logic languages/formalism $\la$. It assumes that, during agent's execution,
the agent can evolve: at each evolution step $i$ the agent's program
(that initially will be $P_0$)
may change (e.g., by learning and via interaction with other agents), with a
transformation of $P_i$ into $P_{i+1}$, thus producing a Program
Evolution Sequence $PE = [P_0 , \ldots , P_n , \ldots ]$. The program
evolution sequence will imply a corresponding Semantic Evolution
Sequence $ME = [M_0 , \ldots , M_n, \ldots ]$ where $M_i$ is the semantic account
of $P_i$ at step $i$ according to the semantics of $\la$.

The agent' \emph{history} $H$, which includes what the agent has recorded of its own activities and of its interaction with the environment, will evolve as well. The history $H$ constitutes in fact the agent's \emph{memory}. We assume $H$ to contain, at least, the set of (the last versions of) \emph{past events}, where past events record the external and internal events that have been perceived (where internal events are those events originated within the agent itself, in the course of its reasoning activities), and the actions that the agent has performed; thus, $H$ defines the up-to-date image that the agent has of its own and of the external world's state of affairs\footnote{For a recent formal approach to memory management in logical agents, cf. \cite{CostantiniP19a,CostantiniP19b}.}. We assume that past events are time-stamped, and that the timestamp is automatically added to newly recorded past events; we omit the explicit indication of timestamps when not needed. When referring to a past event, we will implicitly refer to its most recent version (the one with the newest timestamp), should several versions exist.

The Evolutionary Semantics
$\varepsilon^{\ag}$ of $\ag$ is thus the tuple $\langle H, PE, ME\rangle$,
with $n = \infty$ (i.e., over a potentially infinite evolution). The \emph{snaphot at stage i}, indicated with $\varepsilon^{Ag}_i$, is the tuple
$\langle H_i, P_i, M_i \rangle$

Notice that states, in our case, are not simply intended as time instants.
Rather, they correspond to stages of the agent evolution. Time in this setting is considered
to be local to the agent, where with some sort of \as internal clock'' is able
to time-stamp events and state changes. We borrow from \cite{Manna1991} the following
definition of \emph{timed state sequence}, that we tailor to our setting.

\begin{definition}
	Let $\sigma$ be a (finite or infinite) sequence of states, where the i-th state $e_i$, $e_i \geq 0$,
	is the \emph{semantic snaphots at stage i}, i.e., $\varepsilon^{Ag}_i$,
	of given agent $\ag$. Let $T$ be a corresponding sequence of time instants $t_i$, $t_i \geq 0$.
	A \emph{timed state sequence for agent $\ag$} is the couple $\rho_{\ag} = (\sigma, T)$. Let $\rho_i$ be the i-th state, $i \geq 0$,
	where $\rho_i$ $=$ $\langle e_i, t_i\rangle$ $=$ $\langle \varepsilon^{\ag}_i, t_i\rangle$.
\end{definition}

We in particular consider timed state sequences which are \emph{monotonic}, i.e., if $t_{i+1} > t_i$ then $e_{i+1} \neq e_i$. In fact, there is no point in semantically
considering a static situation: as mentioned, a transition from $e_i$ to $e_{i+1}$ will in fact occur when something
happens, externally or internally, that affects the agent.

Then, in the above definition of A-ILTL operators, it is immediate to
let $s_i = \rho_i$ (with a refinement, cf. \cite{LANMR2012},
to make states correspond to
time instants).

We need to adapt the interpretation function $\ii$ of LTL to our setting. In fact, we intend to employ A-ILTL within
agent-oriented languages, where we restrict ourselves to logic-based languages
for which an evolutionary semantics and a notion of logical consequence can be defined.
Thus, given agent-oriented language $\la$ at hand,
the set $\Sigma$ of propositional letters used to define an A-ILTL semantic framework
will coincide with all
ground expressions of $\la$ (an expression is \emph{ground} if it contains no variables, and
each expression of $\la$ has a possibly infinite number of ground versions).
A given agent program can be taken as standing for its (possibly infinite) ground version, as
it is customarily done in many approaches. Notice that we have to distinguish between logical consequence in $\la$,
that we indicate as $\models_{\la}$, from logical consequence in A-ILTL, indicated above simply as $\models$.
However, the correspondence between the two notions can be quite simply stated by specifying that
in each state $s_i$ the propositional letters implied by the interpretation function $\ii$ correspond to
the logical consequences of agent program $P_i$:

\begin{definition}
	Let $\la$ be a logic language. Let $\mathit{Expr}_{\la}$ be the set of ground expressions that can be built from
	the alphabet of $\la$. Let $\rho_{\ag}$ be a timed state sequence for agent $\ag$, and let
	$\rho_i = \langle \varepsilon^{\ag}_i, t_i\rangle$ be the ith state, with $\varepsilon^{\ag}_i = \langle H_i, P_i, M_i \rangle$.
	An A-ILTL formula $\tau$ is defined over sequence $\rho_{\ag}$ if in its interpretation
	structure ${\cal{M}} = \langle \mathbb{N}, \ii \rangle$, index $i \in \mathbb{N}$ refers to $\rho_{i}$, which means that
	$\Sigma = \mathit{Expr}_{\la}$ and
	$\ii\,:\,\mathbb{N}$\,$\mapsto$\,${{2^{\Sigma}}}$ is defined such that, given $p \in \Sigma$, $p \in \ii(i)$ iff
	$P_i \models_{\la} p$. Such an interpretation structure will be indicated with $\maag$.
	We will thus be consequently able to state whether $\tau$ holds/does not hold w.r.t. $\rho_{\ag}$.
\end{definition}

A-ILTL properties will be verified at run-time,
and thus they can act as \emph{constraints} over the agent behaviour\footnote{By abuse of notation we will
	indifferently talk about A-ILTL rules, expressions, or constraints.}. In an implementation, verification may not 
occur at every state (of a given interval). Rather, sometimes properties
need to be verified with a certain frequency, that can be
specific for each property.
To model a frequency $k$, we introduce a further
extension that consists in defining subsequences of the sequence of all states:
if $Op$ is any of the operators introduced in A-ILTL and $k > 1$,
$Op^k$ is a semantic variation
of $Op$ where the sequence of states $\rho_{\ag}$ of given agent
is replaced by the subsequence
$s_0,s_{k_1},s_{k_2},\ldots$ where for each $k_r, r \geq 1$, $k_r\ mod\ k\ = 0$, i.e., $k_r = g \times k$ for some $g\geq 1$.

A-ILTL formulas to be associated to an agent to establish the properties it has to fulfil can be defined within the
agent program, though they conceptually constitute an additional separate layer. Agent evolution can be considered to be \as satisfactory'', or \as coherent'',
if it obeys all these properties.
An \as ideal'' agent will have a coherent evolution. Instead, violations will occasionally occur,
and actions should be undertaken so as to attempt to regain coherence for the future.

It is important to observe that, A-ILTL expressions are not built-in in any agent program (though some basic ones might be). Rather, they are defined by the agent's designer, according to the application at hand. In fact, in the following sections we will outline many applications of A-ILTL expressions, and some useful extensions to their basic form. Our examples will concern ethics but also other issues: as said in the Introduction, we propose in fact a toolkit for run-time self-checking (and self-correction/improvement, as we will see) which is particularly suitable for ethical control in the sense of \cite{Russel2019}, but can be useful to many purposes.

\section{A-ILTL for Reflexive Self-Checking: Liveness and Safety Properties}\label{lsproperties}

In this section we illustrate the usefulness of
A-ILTL constraints to define and check liveness and safety properties, and to define complex reactive patterns. To this aim, we use
the \emph{pragmatic} form that we have adopted in DALI\footnote{cf. Subsection\rif{dalidescr} below for a short description of the main features of the DALI language.}, where
an A-ILTL expression is represented as $\mathit{OP(m,n;k)}\,\varphi$. Herein,
$m,n$ define the time interval where (or since when, if $n$ is omitted) expression $OP\,\varphi$ is required to
hold, and $k$ (optional) is the frequency for checking whether
the expression actually holds. For instance, $\mathit{EVENTUALLY(m,n;k)}\,\varphi$ states that
$\varphi$ should become true at some point between time instants $m$ and $n$. Notice in fact that A-ILTL constraints act as monitors, where each constraint however is not checked continuously, but rather at a certain frequency, that will be related by a designer to the intended meaning of the constraint itself. A default frequency is provided if $k$ is not specified.

In rule-based logic programming languages like DALI,
we restrict $\varphi$ to be a conjunction
of literals.
In pragmatic A-ILTL formulas, $\varphi$ must be ground
when the formula is checked. However, we allow variables to occur in an A-ILTL
formula, to be instantiated via a \emph{context} $\chi$ (we then talk about {\em contextual A-ILTL formulas}).
Notice that, for the evaluation of $\varphi$ and $\chi$, we rely upon the
procedural semantics of the `host' language.

In the following, a contextual A-ILTL formula $\tau$ will implicitly stand for the ground A-ILTL formula
obtained via evaluating the context. 

The following formulation deals with complex reaction according to a temporal condition.
The way reaction is performed will depend upon the underlying language $\la$, and will be defined
by an expression (a single statement, a sequence of statements, or an entire subprogram) that we call \emph{reactive pattern}. 

\begin{definition}
	A reactive A-ILTL rule is of the form (where $M,N,K$ can be either variables or constants)\\
	\tbl\tbl $\mathit{OP(M,N;K)\varphi :: \chi \div \react}$\\
	where (i) $\mathit{OP(M,N;K)\varphi :: \chi}$ is a contextual A-ILTL formula, called the \emph{monitoring condition},
	that should involve the observation of either external or internal events;
	(ii) $\react$ is called the {\em reactive part} of the rule, and is a reactive pattern.
\end{definition}

Whenever the monitoring condition (automatically checked at frequency $K$)
is violated (i.e., it does not hold) within given interval, then the reactive part $\react$ is executed.

Take for instance the example of a controller that has to keep the temperature in office hours
(say between 8 a.m. and 5 p.m.) in the range 19--21 (celsius degrees). In this case,
$\mathit{temperature_N}$ is a \emph{present event} ($N$ standing for \emph{now}), i.e., the current value of an external event which is observed at a certain frequency by
the system. If the condition is violated, a reaction will try to restore the wished-for
situation. However, we assume to be in a smart building (where in fact the temperature is
monitored by intelligent agents) where the agent is able to select, in order to modify the
temperature, the best suitable energy source, for instance the less expensive.
Notice that at different times of the day different sources of energy can be less expensive.
Remember also that the A-ILTL constraint is dynamically checked at a certain
frequency, here ten minutes (which will be the default one if no frequency is specified explicitly).
So, in a given interval the monitoring condition will sometimes succeed (then nothing is done)
and sometimes fail. In the latter case, the font of energy $S$ which is cheaper
\emph{in that moment} is used in order to suitably affect the temperature and try to keep it
in the specified range.
In the proposed approach, this can be formalized as follows (where, as there are no variables,
context is omitted, and $\mathit{modify\_temperature_A}$ is an action). The expression that allows $S$ to be selected within a set of alternatives according to some kind of preference (here on cost), can be expressed in any of the existing preference mechanisms for logic programming languages (cf., e.g., \cite{CostantiniF11,CostantiniF12,CF-jalgor09}).

\medskip
$
\begin{array}{ll}
	\mathit{ALWAYS(8:00\,a.m.,\,5:00\,p.m.;10m)}\\
	\tbs 19 \leq \mathit{temperature_N} \leq 21\ \div\\
	\tbs \mathit{modify\_temperature_A(S), S\ IN}\\
	\tbl \{\mathit{external\_electricity},\\
	\tbl\,\ \mathit{gas},\\
	\tbl\,\ \mathit{solar\_panel\_electricity}\,:\\
	\tbl\tbl\tbl\mathit{less\_expensive}\}
\end{array}
$

The next example is a meta-statement expressing single-minded commitment
in agents, i.e., that a goal should be pursued until reached, or no longer believed possible. In this example, the constraint performs an act of introspection to access and evaluate aspects of the agent's BDI state. This requires that such aspects are suitably represented at the metalevel.
The fact that a goal $G$ is possible is evaluated, in our formalization, w.r.t. a module $M$
that represents the context for $G$, via a `possibility' predicate $P$: so, $G$ is deemed to be possible w.r.t. $M$ if $P(G,M)$ is true. How to define such a predicate is discussed in \cite{Cos2012ASPmodules}, where the choice is to represent and evaluate $M$ as an Answer Set Programming (ASP) module; here, $M$ should be such a module.
In case the goal is still deemed to be possible and is not timed-out but has not been achieved
yet, then the reaction consists in re-trying the goal, which is an action that might imply either resuming
a suspended plan, or a re-planning.

\medskip
$
\begin{array}{ll}
	\mathit{NEVER}\\
	\tbs \mathit{goal(G)},\\
	\tbs \mathit{eval\_context(G,M),P(G,M)},\\
	\tbs \mathit{not\ timed\_out(G)}\\
	\tbs \mathit{not\ achieved(G)} \div\\
	\tbl \mathit{retry_A(G)}
\end{array}
$

Another possibility is not simply retrying the goal, but also
reconsidering the evaluation context, that might for some reason
have become obsolete. Thus, the reactive part might be
\[\mathit{reconsider\_context(G,M,M'),P(G,M'),retry_A(G)}\]
here, the module for evaluating possibility could be updated, and this might lead to either continue or stop retrying the goal.

Each element of the conjunction composing the reactive part can have preconditions.
If preconditions do not hold for some element, then that element is skipped.
One could for instance add the precondition that a goal
can retried if sufficient resources are available, i.e.,
\[\mathit{retry_A(G)} \PRE \mathit{have\_resources(G)}\]
where the goal would not be retried in the negative case.

The following expression states that any goal $G$ that the agent may have formed due to its interaction with the environment has to be dropped if not coherent with designer's intention or user's interests. This is very important, because, as discussed by Stuart Russel in his recent book \cite{Russel2019}, agents that learn can dangerously depart from the behaviour that is expected from them. So, conformity of an agent's goals to specification, or however adherence to user approval, must be constantly verified. Here, the frequency-based checks and the introspective capabilities of A-ILTL constraints play a relevant role, as they are able to detect changes in an agent `mental state' that may happen over time in an unpredictable way.

\medskip
$
\begin{array}{ll}
	\mathit{ALWAYS}\\
	\tbs \mathit{goal(G)},\\
	\tbs \mathit{not\ designer\_specified(G)}\ \mathit{OR}\\
	\tbs \mathit{not\ user\_approved(G)}\ \div\\
	\tbl \mathit{drop(G)}
\end{array}
$

Notice that in the examples we used some metapredicates which are reminiscent of the BDI model, i.e., $\mathit{goal(G)}$,
$\mathit{timed\_out(G)}$, $\mathit{achieved/not\ achieved(G)}$,
$\mathit{retry(G)}$. Such predicates explicitly represent (we might say \emph{reify}) elements of the agent's operation, so that such elements can be evaluated and, possibly, affected: in the last example, the A-ILTL constraint may decide that a goal can be dropped. According to the `host' language, such predicates might be pre-defined, or they might be fully user-defined. For instance in DALI they are, at the moment, user-defined, so for instance how to assume or drop a goal must be determined by a piece of code written by a programmer. The possibility of making (at least some of them) pre-defined, and which mechanisms to implement to affect the agent's internal operation is under careful consideration. 

\section{Evolutionary Expressions}
\label{beh-rul}

It can be useful to define properties
to be checked upon arrival of sequences of events, of which
however only relevant ones (and their order) should be considered.
To this aim, we introduce a new kind of A-ILTL constraints,
that we call \emph{Evolutionary A-ILTL Expressions}.
To define partially known sequences of any length,
we admit for event sequences a syntax inspired to that of regular expressions
so as to be able to ignore irrelevant/unknown events, and repetitions (cf. \cite{LANMR2012}). Notice that, the incoming event sequence is represented as a sequence of past events, ordered by their timestamp (that we omit when not needed).

\begin{definition}[Evolutionary A-ILTL Expressions]
	\label{evexps}
	Let $\calsevp$ be a sequence of past events, and $\calsf$ and $\calsj$ be sequences of events.
	Let $\tau$ be a contextual A-ILTL formula $\op\,\varphi\,::\,\chi$.
	
	\noindent
	An \emph{Evolutionary A-LTL Expression} $\varpi$ is of the form \[\calsevp\,:\,\tau\,:::\,\calsf\,::::\,\calsj\]
	where:
	(i) $\calsevp$
	denotes the sequence of relevant events which are supposed to have happened, and in which order,
	for `triggering' the rule; i.e.,
	this event sequence acts as precondition: whenever one or more of these events happen (and are thus recorded) in the specified order, $\tau$ will be checked (i.e., check of $\tau$ is triggered by any prefix of $\calsevp$);
	(ii) $\calsf$ denotes the events that are expected to happen in the future
	without affecting $\tau$;
	(iii) $\calsj$ (optional) denotes the events that are expected \emph{not} to happen in the future;
	i.e., whenever any of them should happen, $\varphi$ is not required to hold any longer, as these are \as breaking events''.
\end{definition}

An Evolutionary A-ILTL Expression can be evaluated w.r.t. a state $s_i$ which includes among its components
the agent's \emph{history}. Precisely, in a state $s_i$, the component $H_i$ of $s_i$ \emph{satisfies}
an event sequence $S$ whenever either no event in $S$ is present in $H_i$, or events are present in $H_i$ which constitute a prefix of $S$, as they occur (according to the timestamps) in the order specified by $S$ itself. All past events (which include past actions performed by the agent) are assumed to be stored in a ground form, and are indicated by the postfix `P' (for instance, in the example below $\mathit{supply_P}$ is a past event).

All variables occurring
in evolutionary A-ILTL expressions are implicitly universally quantified, in the style of Prolog-like logic languages. The context can be omitted if not needed.

A sample evolutionary A-ILTL expression is the following (where $N$ stands for
operator \as never''):

\meno
$
\begin{array}{ll}
	\mathit{supply_P}^{+}(r,\_s)\,:\, N(\mathit{quantity}(r,V), V < th)\,:::\ \ 
	\mathit{consume_A}^{+}(\_r,Q)
\end{array}
$

\medskip
Syntactically: $\mathit{supply_P}^{+}(r,\_s)$ stands for a sequence (of unknown length) of \emph{past} supply actions of unknown quantities (`unknown value' $\_s$), performed by the agent itself or by some other agent, with the effect to replenish the agent's stock or resource $r$; $\mathit{consume_A}^{+}(\_r,Q)$ stands for a sequence (of unknown length) of \emph{future} consumption actions of certain quantities (`unknown value' $\_$r) of resource $r$, that the agent may perform. 
The `core' A-ILTL expression $N(\mathit{quantity}(r,V), V < th)$, $N$ standing for `never', specifies that, whatever the supply and consumption, the available amount $V$ of the resource
$r$ must remain over a certain threshold $th$, i.e., $V$ should \emph{never} be less than $th$. 

Such expression is supposed to be checked at run-time at a certain frequency (here the default one, not having the frequency been specified explicitly) whenever a supply action is performed (and thus recorded), which makes the precondition verified.
A violation may occur if in some state the A-ILTL formula $\tau$ does not hold, i.e., in the example, if the available quantity $V$ of resource $r$ runs too low. Notice that, since the constraint is checked periodically, it might be the case that the condition $\mathit{quantity}(r,V), V < th$ be momentarily violated between one check and the subsequent one. To avoid possible misfunctionings deriving from this problem, either the frequency must be suitably increased, or the quantity $V$ must be set to a precautionary higher value, in order not to really arrive at a too low value.

The above constraint is significant from an ethical point of view: in fact, a very common unethical behaviour concerns the improper use/waste of limited resources. Think, for instance, of the excessive and/or improper use of environmental resources, like, e.g., water.

\medskip
Below, we formally state when an Evolutionary A-ILTL Expression holds or not.

\begin{definition}
	An Evolutionary A-ILTL Expression ${\varpi}$, of the form specified in Definition\rif{evexps}:
	\begin{enumerate}
	\item
	\emph{holds} in state $s_i$ whenever (i) history $H_i$ satisfies $\calsevp$ and $\calsf$ and does not include any event in $\calsj$, and $\tau$ holds or
	(ii) $H_i$ includes some event occurring in $\calsj$ (we say that the expression is \emph{broken}); 
	\item 
	is \emph{violated} in state $s_i$ whenever
	$H_i$ satisfies $\calsevp$ and $\calsf$ and does not include any event in $\calsj$,
	and $\tau$ does not hold.
\end{enumerate}
\end{definition}

Operationally, an Evolutionary A-ILTL Expression can be finally deemed to hold if
either the upper bound of the specified interval has been reached (if a finite interval has been specified) and $\tau$ holds, or an unwanted event has occurred.
Instead, an expression can be deemed \emph{not} to hold (or, as we say,
to be \emph{violated} as far as it expresses a wished-for property) whenever
$\tau$ is false at some point without the occurrence of breaking events.
In this case, a repair action (like in reactive A-ILTL rules) can be optionally specified.

For instance, in the variation of previous example listed below a repair measure is specified, which 
states that no more consumption can take place if
the available quantity of resource $r$ is scarce.

\meno
$
\begin{array}{ll}
	\mathit{supply_P}^{+}(r,\_s)\,:\, N(\mathit{quantity}(r,V), V < th)\,:::\\ 
	\hspace{3cm}\mathit{consume_A}^{+}(r,Q)\,\div \mathit{block}(\mathit{consume_A}(r,Q))
\end{array}
$

\medskip
We might instead opt for another (softer) formulation, that forces the agent to limit consumption to small
quantities (say less than some quantity $q$) if the threshold is approaching (say that the remaining quantity is $th + f$, for some $f$).

\meno
$
\begin{array}{ll}
	\mathit{supply_P}^{+}(r,\_s)\,:\, N(\mathit{quantity}(r,V), V < th + f)\,:::\\ \hspace{3cm}\mathit{consume_A}^{+}(r,Q)\,\div 
	\mathit{allow}(\mathit{consume_A}(r,Q), Q < q)
\end{array}
$

\medskip
The above example demonstrates that the proposed
approach to dynamic verification is indeed needed: any sequence of performed `supply' and `consume'
actions may arrive, so the number of potential configurations is not limited and static verification methods appear hardly applicable. One might provide total
supply and consumption figures: however, one would just draw the quite pleonastic conclusion that the desired property
holds whenever supply is sufficiently generous and consumption prudentially limited. Instead, in the proposed approach
we are able to verify the target property \as on the fly'', whatever the sequence of performed actions
and the involved quantities. Moreover, we are also able to try to
repair an unwanted situation and regain a satisfactory state of affairs.

Below we provide another example of an Evolutionary A-ILTL expression that, though simple, is in our opinion significant as it is
representative of many others. Namely, we assume that an agent manages
a FIFO queue, thus accepting operations of pushing and popping elements on/from the queue.
The example thus models in an abstract way
the very general and widely used architecture where an agent provides
a service to a number of `consumers'. 
We establish the restriction
that the queue must never contain any duplicated elements $e_1$ and $e_2$. 
This means that customers cannot reiterate a
request of service if their previous one have not been processed. This 
in order to ensure fairness in the satisfaction of different customers' requests.
The possible actions are: $\mathit{push_A}(Req,Q)$, that is performed by other agents and inserts a generic value $Req$ in the queue, representing (in some format) a request of service (each inserted element is given an index $e_i$);
$\mathit{pop_A}(e,R)$, that extracts the oldest element from the queue, i.e., the request to be presently processed.
The A-ILTL expression  considers an unknown number of pushing actions happened in the past (and thus are
now recorded as past events) and can foresee an unknown number of future popping actions.

\meno
$
\begin{array}{ll}
	\mathit{push_P}^{+}(Req,Q)\,:
	\hspace{0.1cm}N(\mathit{in\_queue(e_1,RX), in\_queue(e_2,RX)})\,:::\,\mathit{pop_A}^{+}(e,Q)
\end{array}
$

\smallskip
This expression will be be the subject of the experiments illustrated in the next section.

The next one is an example of Evolutionary A-ILTL Expression that might occur in an agent program installed on an autonomous robot working on batteries, which is able to check its own charge level. The robot moves in some environment 
to perform some task, thus consuming battery. The following A-ILTL axiom states that, after a battery recharge (indicated as a past event, postfix '$P$') at time $T$, the charge level should be sufficient for six hours despite a sequence of actions which can be considered to be `normal' in relation to the robot's task. These actions may for instance involve moving around, cleaning rubbish, delivering packages, etc. Instead, the charge level can be expected to be low (the property is `broken') in case of extensive usage actions, for instance in case of an exceptional unexpected event that requires the robot to increase its activities (e.g., drying water in case of a flooding from a broken pipe). There must be of course a classification, in the agent's background knowledge base, of what should be intended by `normal' or `extensive' usage. 

\medskip
$
\begin{array}{ll}
	\mathit{recharge\_battery_P}\!:\! T\,:\\
	\tbs \mathit{ALWAYS}(T,T+6_{hour})\ \,charge\_level(L), L > \mathit{low}\\
	\tbs:::\mathit{normal\_usage\_action(Act)*}\ ::::\mathit{extensive\_usage\_action(Act)*}
\end{array}
$

\medskip
The above expression should be combined with another A-ILTL expression, seen below, which forces recharge every six hours.
This one should state that if the last battery recharge $\mathit{recharge\_battery_P}$ has occurred at time $T$ 
which is more than six hours different from present time $\mathit{now}$, then
the goal $\mathit{recharge\_battery_G}$ must be set (where postfix `G' stands for `goal'). Achieving this goal may require, for instance, reaching the nearest recharge station. 

\medskip
$
\begin{array}{ll}
	\mathit{ALWAYS}\\
	\tbs \mathit{recharge\_battery_P}\!:\!\mathit{T, now - T > 6_{hour}}\ \div\ \mathit{recharge\_battery_G}
\end{array}
$

\medskip
Whenever an Evolutionary A-ILTL expression is either violated or broken, not only an immediate reaction can be attempted, but measures can be undertaken aimed
at recovering a desirable or at least acceptable agent's state.
\begin{definition}
	An evolutionary LTL expression with repair $\varpi^{r}$ is of the form
	${\varpi} | \repair_1 || \repair_2$
	where ${\varpi}$ is an Evolutionary LTL Expression adopted in language $\la$, and $\repair_1, \repair_2$ are atoms of $\la$.
	$\repair_1$ will be executed
	(according to $\la$'s procedural semantics) whenever $\varpi$ is violated, and $\repair_2$ will be executed whenever $\varpi$ is broken.
	$\repair_1$ and $\repair_2$ are called \emph{countermeasures}.
\end{definition}

In the robot example, whenever a low level of charge is detected,
the immediate reaction can be to stop the robot's operation.  However, there can be the case of low battery under normal usage, that might imply a fault either in the battery or in the recharge station. Countermeasure $\repair_1$ in fact, may (for the sake of the example) alert the user. Instead $\repair_2$, taken in case of low battery under exceptional usage, will simply imply the robot to resort to the recharge station.
The overall expression will take the form:

\medskip
$
\begin{array}{ll}
	\mathit{recharge\_battery_P}\!:\! T\,:\\
	\tbs \mathit{ALWAYS}(T,T+6_{hour})\ \,charge\_level(L), L > \mathit{low}\\
		\tbs:::\mathit{normal\_usage\_action(Act)*}\ ::::\mathit{extensive\_usage\_action(Act)*}\\
		\tbs\ \div \mathit{stop\_robot\_operation}\\
	\tbs\ \ |\ \mathit{alert\_user\_possible\_fault_A}\ ||\ \mathit{recharge\_battery_G}
\end{array}
$

\medskip
Evolutionary A-ILTL expressions can be further enhanced (by means of a slight extension to the above definition) by introducing
a third kind of counter-measure, aimed at preventing a potentially breaking event from
disrupting the wished-for property. In the following example, there has been an accident at place $D$ at time $T$, and an ambulance has been sent for rescue. The condition is that the rescue should never arrive late. However, there is news of a traffic jam that blocks the ambulance. In the example, the new kind of counter-measure consists in sending either an helicopter 
or a coast guard boat, with preference to the option which is evaluated as 
more effective in terms of time for reaching place $D$ from the rescuers' present location\footnote{The construct used to express the preference has been discussed and formalized in \cite{CF-jalgor09,CostantiniF11,CostantiniF12}}.

\medskip
$
\begin{array}{ll}
	\mathit{accident_P(D)}\!:\! T\,:\\
	\tbs \mathit{NEVER\ \,late\_rescue(D,T)}\\
	\tbs::::\mathit{traffic_P,\,ambulance\_sent_P,\,ambulance\_blocked_P}\ \div\\
	\tbs\ \ |||\ \mathit{alternative\_transportation\ IN}\\
	\tbl\tbs\tbs\{\mathit{elicopter,boat : faster\_reach(here,D)}\}
\end{array}
$

\section{Experimental Evaluation}
\label{experiments}

We have implemented the proposed approach
within the DALI multi-agent system \cite{CostantiniGPS17}.
In this section, we present some experiments, aimed to establish
the effectiveness of the approach. In particular, we wish to practically demonstrate that the use of A-ILTL expressions is computationally affordable, in the sense that they do not slow down an agent’s operation, while, on the contrary, using them is even more efficient than using ad-hoc solutions. 

We could not establish a comparison w.r.t. competitor approaches, that at present do not exist. So, we compared our approach w.r.t. a correspondent solution developed in pure Prolog\footnote{The author wishes to thank former Ph.D. student Dr. Alessio Paolucci who has written the code and practically performed the experiments.}. Notice that, DALI
is in fact an agent-oriented extension to Prolog whose interpreter is implemented in Prolog itself so, when stripped of its peculiar features, DALI \as collapses'' into Prolog. 

Below, we first preliminarily briefly illustrate the DALI language \cite{jelia02,jelia04} in order to make a reader able to understand the code. Then, we show the alternative (Prolog and DALI) solutions, and finally we propose their experimental comparison. 

The experiments concern 
the queue constraint illustrated above:

\meno
$
\begin{array}{ll}
	\mathit{push_P}^{+}(Req,Q)\,:
	\hspace{0.1cm}N(\mathit{in\_queue(e_1,RX), in\_queue(e_2,RX)})\,:::\,\mathit{pop_A}^{+}(e,Q)
\end{array}
$

\medskip
here, a Queue agent $Q$ considers an unknown number of pushing actions happened in the past (recorded as past actions, postfix $P$) and expects an unknown number of future popping actions, each one always returning the \as oldest'' element of the queue in response, whereas the agent keeps checking that the queue never contains duplicated elements. 

\subsection{DALI in a Nutshell}
\label{dalidescr}

In DALI,
the autonomous behaviour of an agent is triggered by several kinds of events, which are \as first-class objects'' in the language syntax and semantics: external
events, internal, present and past events.

{\bf External events} are
syntactically indicated by the postfix \textit{E}. Reaction to each such event is defined by a
reactive rule, denoted by the
special token $:>$.
The agent remembers to have reacted by converting an
external event into a \textit{past event} (postfix \textit{P}).
An event perceived but not yet reacted to is
called \as present event'' and is indicated by the postfix \textit{N}. It is often useful for an agent to reason about present events, that make the agent aware of what is happening in its external environment.

{\bf Actions} (indicated with postfix \textit{A}) to be performed by DALI agents may
have or not have preconditions: in the former case, the actions are
defined by \as actions rules'', in the latter case they are just action
atoms. The new token $:<$ characterizes an action rule that specifies an action's preconditions.
Similarly to events, actions are recorded as
past actions.

{\bf Internal events} is the device which
makes a DALI agent proactive.
An internal event is syntactically indicated by the
postfix \textit{I}, and its description is composed of two rules.
The first one contains the conditions (knowledge, past events,
past actions, etc.) that must hold so that the reaction (in the
second rule) is triggered. Thus, a DALI agent is able to react to its own conclusions, therefore enacting \as spontaneous'' proactive behaviour, i.e., behaviour not directly dependent upon external stimuli.
Internal events are automatically attempted at a default
frequency, customizable by user directives.

Agents usually record events that happened and actions
that they performed. Notice in fact that an agent can describe the state of the world only in terms
of its perceptions, where more recent remembrances define the agent's approximation
of the current state of affairs. We thus define set ${\cal P}$ of current (i.e., most recent) past
events and actions (each one time-stamped), and a set $PNV$ where we store previous ones (where a designer can specify which past events to keep and which to cancel, and under which conditions).

The DALI communication architecture \cite{dali-communication} implements the DALI/FIPA protocol, which consists of the main FIPA primitives\footnote{FIPA is a widely used standardized ACL (Agent Communication Language), cf. \url{http://www.fipa.org/specs/fipa00037/SC00037J.html} for language specification, syntax and semantics.}, plus few new primitives which are peculiar to
DALI. Each DALI agent has its own customizable filter for incoming and outgoing messages, composed by user-definable metarules which are to be specified in a special file. So, a message will then be sent/received if the metarule rule for the primitive used is present in the communication file, and the conditions are met. It is also possible not to enter conditions, but to use 'true' instead, which implies that the message will always pass. 
In addition, there are rules for meta-reasoning which allow the agent to consult its knowledge and ontologies for understanding incoming messages.
Notice that, DALI has been made compatible with the Docker technology (cf. \cite{CostantiniGPS17} for details). So, a DALI agent can be deployed within a container.

The semantics of DALI is based upon the declarative semantic framework
introduced in \cite{postdalt06}. 
DALI has been fully implemented on the basis of Sicstus Prolog \cite{carlsson2010sicstus}, and the DALI programming environment is freely available at \small\url{https://github.com/AAAI-DISIM-UnivAQ/DALI}\normalsize. The DALI framework has been experimented and practically applied in many, also industrial, applications.

DALI is a general-purpose agent-oriented programming language, non-committal w.r.t. any agent architecture. However, it has features that can emulate a BDI-oriented language such as AgentSpeak. In particular, DALI provides {\bf Goals},
syntactically indicated by the postfix \textit{G}, which are special internal events that, when triggered, are executed only once (i.e., they are not attempted periodically). This construct emulates AgentSpeak's \emph{plans}, as the first rule provides the context that, if verified, triggers the execution of the second rule (where in AgentSpeak these two components are joined within a unique rule). Moreover, DALI is equipped with a plugin to an ASP solver: this allows an agent to compute entire plans that can then be inspected, evaluated, executed, re-evaluated, etc., in the BDI fashion, according to the desired level of commitment of the agent to the current goal.

For exploring advanced DALI features such as the communication architecture,
the integration with the Docker technology, the web interface, the cloud implementation, the ability to use Redis as a database, cache, message broker, and interface with Phyton, and the interoperability with agents written in other languages, plus the integration with ASP, and more, the reader may refer to \cite{dali-communication,CostantiniGPS17,CDGSP17,PADL2017} and to the github repository.

\subsection{Pure Prolog Code}
\label{prolog-code}

Below we report the code of the version of the Queue agent implemented in DALI, where however the specific DALI features are employed only for the program activation via an external event triggering a reactive rule. The rest of the program is instead written using
Prolog only. The test agent gets active and performs a test session
whenever it receives from the user a message with content $run\_pure\_test(Times)$
where $Times$ specifies the number of elements to be pushed and popped on the queue.

When the agent receive the event $run\_pure\_test$ it reacts, thus invoking the
$run\_pure\_testing$ subgoal\footnote{the term `subgoal' is meant here and below in the Prolog sense, as a `procedure` to execute or, logically, an atom to prove, with no relation to the BDI meaning.} with $Times$ as parameter.
$run\_pure\_testing$ prints information for the user on the console, and
starts the 'pushing' phase.
The $pushing(Times)$ goal repeatedly pushes an item (through $push\_item$ subgoal),
as far as $Times > 0$, and then it ends. To do so it makes use of recursion.

$push\_item$ is responsible of items pushing and, as first step,
retrieves the data structure $pqueue(Q)$.
Then it randomly generates an item, and checks if that item
is already present in the queue. If it alreay exists,
then $push\_item$ fails, and this item pushing is skipped. This implements the `NEVER' condition in the A-ILTL constraint.
If the new item is not in the queue, then it is added in the head. The old queue is
removed from memory ($retract(pqueue(Q))$), and the new one is pushed
into the knowledge base ($assert(pqueue(NQ))$.
$popping$ is then invoked to perform items removal from the queue.
It makes use of $pop\_item$, a subgoal that retrieves and unifies the
queue through $clause(pqueue(Q),\_)$, and then extracts the head of the
queue $Q$. After the 'popping' phase, the test ends. The time spent in performing the test is displayed.

\medskip\noindent $
\begin{array}{ll}
	run\_pure\_testE(Times)\THEN & run\_pure\_testing(Times).\\
	\\
	run\_pure\_testing(Times)\IF & pretty\_start,\\
	& now(StartTime),\\
	& T1\ is\ Times + 1,\\
	& nl, write('PUSHING...'), nl,\\
	& pushing(T1),\\
	& nl, write('POPPING...'), nl,\\
	& popping(T1),\\
	& now(EndTime),\\
	& TestTiming\ is\ EndTime - StartTime,\\
	& nl, write('Time: '), write(TestTiming), nl,\\
	& pretty\_end.
\end{array}
$

\medskip
\medskip\noindent $
\begin{array}{ll}
	pushing(0).\\
	pushing(Times)\IF &push\_item, T1\ is\ Times - 1, pushing(T1).\\
	\\
	push\_item\IF & clause(pqueue(Q),\_),\\
	& random(1,300,Item),\\
	& not(exists\_in\_queue(Item,Q)),\\
	& nl, write('Pushing: '), write(Item),\\
	& append(Q, [Item], NQ),\\
	& retract(pqueue(Q)),\\
	& assert(pqueue(NQ)),\\
	& nl, write('Queue: '), write(NQ), nl.
\end{array}
$

\medskip
\medskip\noindent $
\begin{array}{ll}
	push\_item \IF & assert(pqueue([])).\\
	\\
	exists\_in\_queue(X,[X|\_])\IF & true.\\
	exists\_in\_queue(X,[\_|Tail])\IF & exists\_in\_queue(X, Tail).
\end{array}
$

\medskip
\medskip\noindent $
\begin{array}{ll}
	popping(0).\\
	popping(Times)\IF & pop\_item, T1\ is\ Times - 1, popping(T1).\\
	\\
	pop\_item \IF & clause(pqueue(Q),\_),\\
	& nl, write('Popping: '),\\
	& Q = [H|T],\\
	& write(H),\\
	& retract(pqueue(Q)), assert(pqueue(T)),\\
	& nl, write('Queue: '), write(T), nl.\\
	\\	
	pop\_item\IF & assert(pqueue([])).\\
	\\
	pretty\_start\IF & nl, write('Start\ testing... '), nl.\\
	pretty\_end\IF & nl, write('Test\ finished... '), nl.
\end{array}
$

\subsection{Proper DALI Code}
\label{dali-code}

The proper DALI implementation makes use of DALI advanced features, in particular exploits
actions, and the ability to remember what happened in the past (past actions).
Basically, the DALI infrastructure makes us able to write the program in a very
comfortable manner: each pushing is an action, so every time the action is performed
in the present, the DALI engine records, for future usage, this action as a past
event. In this way, we very simply simulate a queue without using lists, asserts, retracts, etc. When a pop needs to be performed on the queue, we use DALI past events,
to remember about actions performed, and so, to retrieve the correct item from the head of the queue,
in a very elegant manner.
The DALI implementation allows us to concentrate on the problem, without focusing that
much on the data structure, in a declarative fashion.

Below we report the code of the advanced version of the Queue agent implemented
in DALI, taking profit of all DALI features. The test agents gets active and performs a test session
whenever it receives from the user a message with content $run\_dali\_test(Times)$
where $Times$ specifies the number of elements to be pushed and popped on the queue.

\medskip\noindent $
\begin{array}{ll}
	run\_dali\_testE(Times) \THEN & dali\_test\_startA, run\_dali\_testing(Times).\\
	\\
	run\_dali\_testing(Times) \WH & dali\_test\_startP.\\
	run\_dali\_testing(Times)\IF & dali\_start\_pushingA,\\
	& dali\_pushing(Times),\\
	& dali\_end\_pushingA,\\
	& dali\_start\_popA,\\
	& dali\_popping(Times),\\
	& dali\_end\_popA,\\
	& dali\_end\_testingA.
\end{array}
$

\medskip
\medskip\noindent $
\begin{array}{ll}
	dali\_pushing(0)\IF &true.\\
	dali\_pushing(Times)\IF & dali\_remember(Times), T1\ is\ Times - 1, dali\_pushing(T1).\\
	\\
	dali\_remember(Times)\IF & random(1,300,Item),\\
	& not(clause(do\_action(dali\_push\_queue(Item,\_), \_) , \_)),\\
	& get\_push\_index(PI),\\
	& dali\_push\_queueA(Item, PI).\\
	\\
	get\_push\_index(I1)\IF & clause(push\_index(Index),\_),\\
	& I1\ is\ Index + 1,\\
	& retract(push\_index(Index)),\\
	& assert(push\_index(I1)).\\
	\\					
	get\_push\_index(Index)\IF & assert(push\_index(1)),\\
	& Index = 1.
\end{array}
$

\medskip
\medskip\noindent $
\begin{array}{ll}
	get\_pop\_index(I1)\IF & clause(pop\_index(Index),\_),\\
	& I1\ is\ Index + 1,\\
	&  retract(pop\_index(Index)),\\
	& assert(pop\_index(I1)).\\
	\\				
	get\_pop\_index(Index)\IF & assert(pop\_index(1)),\\
	& Index = 1.
	\\
	dali\_popping(0)\IF & true.\\
	dali\_popping(Times)\IF & dali\_forget(Times), V1\ is\ Times, -\, 1, dali\_popping(V1).\\
	\\
	dali\_forget(Dummy)\IF & get\_pop\_index(Index),\\
	& clause(do\_action(dali\_push\_queue(Item,Index), \_) , \_),\\
	& dali\_pop\_queueA(Item).
\end{array}
$

\subsection{Experiments}

The experiments have been performed on a Microsoft Surface Pro 7 PC, equipped with Intel(R) Core(TM) i7-1065G7 CPU@1.30GHz-1.50GHz with 16Gb RAM, using Sicstus 4.6 as the Prolog interpreter.

We did not consider the frequency of constraint-checking, which is available in DALI, but
could not be implemented in an acceptably simple way in Prolog. So, this feature alone constitutes a significant enhancement of DALI with respect to Prolog. 

\begin{figure}
	\centering
	\includegraphics[scale=0.45]{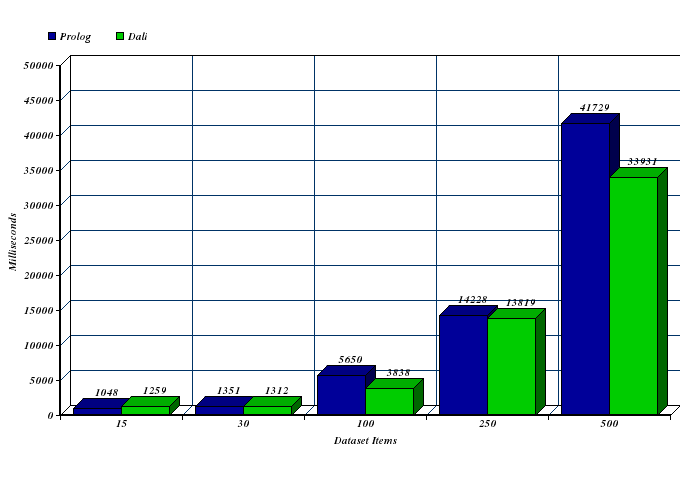}
	\caption{x axis: instance size; y axis: execution time, blue bars pure Prolog green bars DALI}
	\label{table1}
\end{figure}

The instance size (number of elements to push and pop on the queue) can be
established by the user when starting a test session. The items to pop/push are, in the experiments,
randomly-generated numbers. 
In Figure\rif{table1} and Figure\rif{table2} we show the execution time of the two solutions at the
growth of the instance size. In Figure\rif{table3} we show the difference in percentage between the
execution times.

\begin{figure}
	\centering
	\includegraphics[scale=0.45]{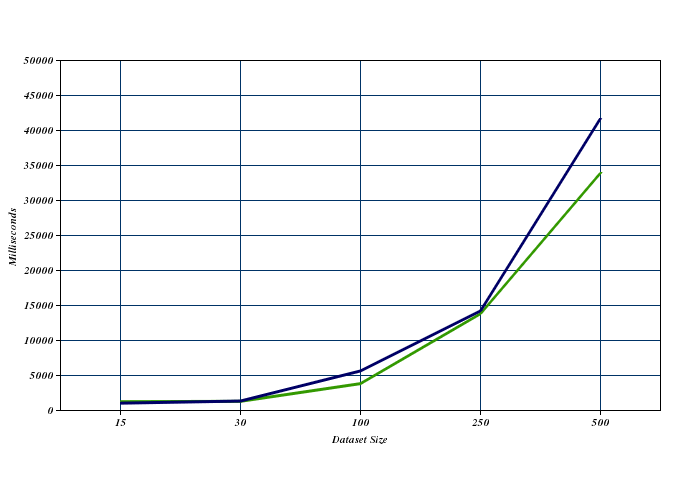}
	\caption{Interpolation average values, blue line pure Prolog green line DALI}
	\label{table2}
\end{figure}

All figures refer to a dataset of up to 500 elements to push and pop. This has been sufficient
to identify an initial \as unstable'' stage and then a trend that further consolidates with the growth of the instance size.

\begin{figure}
		\centering
	\includegraphics[scale=0.45]{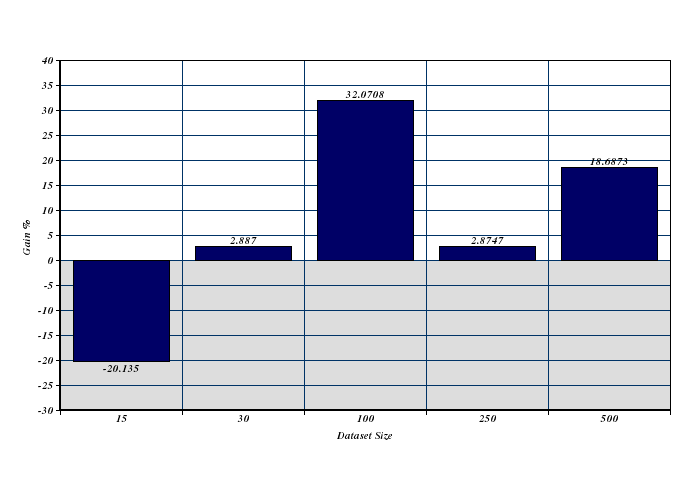}
	\caption{x axis: instance size; y axis: gain (in percentage) when using DALI}
	\label{table3}
\end{figure}

What we can see is that, when the number of events
that we consider is small, then the two solutions are more or less equivalent, the
Prolog one a bit better as it involves no overhead (while the DALI events and actions management necessarily involves some). But, as soon as the instance size grows, the DALI solution becomes
more performant, despite the overhead of the DALI implementation.
We can thus conclude that the new constructs that we propose are not only expressive
and then useful for specifying problem features in a compact and declarative way, but they also
improve efficiency and thus effectiveness of solutions.

\section{Complexity of Check and Discussion}
\label{complexity}

In this section we present an analysis of the complexity of checking A-ILTL expressions.
Let us make the simplifying assumption
that all expressions are checked at the same frequency: i.e., the agent devotes 
(with a certain periodicity) some amount time to perform the checks. Here we try to evaluate 
this amount. Let us assume to have $f$ A-ILTL expressions, and that the time for retrieving
each expression from the computer memory is $m$. Thus, retrieving all expressions to be evaluated
is ${\cal O}(f \star m)$. Let $k$ be the number of the different A-ILTL operator occurring in
the $f$ expressions. Let $\mathit{if\_eval}$ be the time needed in order to understand whether each expression
needs to be evaluated in the present state: this includes checking the related time interval and,
in case of Evolutionary A-ILTL Expressions,
checking the event sequence $\calsevp$ w.r.t. current agent's history. Let $\mathit{max\_eval}$ be the maximum time
needed for the evaluation of each contextual A-ILTL formula $\op\,\varphi\,::\,\chi$.
Let $\mathit{if\_viol\_or\_broken}$ be the maximum time needed to state whether each Evolutionary A-ILTL Expressions
is either violated or broken:
this implies checking event sequences $\calsf$ and $\calsj$ w.r.t. current agent's history.

Therefore, the total time to be spent for checking all A-ILTL Expression (in the worst
case, where all of them are of the Evolutionary kind, and each of them needs to be evaluated at
the present state) can be estimated to:

\[{\cal O}((f \star m ) + (f \star (\mathit{if\_eval} + \mathit{max\_eval} + \mathit{if\_viol\_or\_broken})))\]

Then, for each expression which is either violated or broken, there will be a time spent
in the recovery and countermeasure actions.

The relatively low complexity of check is due to the definition of A-ILTL in relation
to the Evolutionary semantics: in fact, it is not needed to implement a temporal logic 
inference engine. Rather, a system will periodically check $\op\,\varphi\,::\,\chi$.
This in the case of simple non-nested A-ILTL expressions. Introducing more
complex expressions is a subject of future work. In practice however, this complexity anyway requires to keep
the number of A-ILTL expressions as low as possible, and to tune frequency carefully,
according to the environment change rate. In fact, despite being useful, sometimes even essential, for a good functioning of a system, dynamic verification may cause a decay in its performances. 

The motivation why, despite the availability of many techniques, the proposed
approach to dynamic verification actually constitutes a step ahead, has been discussed concerning the above example of supply and consumption.
Nonetheless, an important topic little considered so far, which
we intend to to tackle in already-planned future work, is that of better identifying the boundary between those properties of a MAS than can be verified statically and the ones which necessarily require dynamic verification. It would be important to shift ahead this boundary, thus simplifying the task of dynamic verification. 

\section{Other Related Works and Concluding Remarks}

\label{conclusions}

In this paper, we have extended past work, so as to devise a toolkit for run-time self-checking of logic-based agents. 
The proposed toolkit is able to detect and correct behavioural anomalies by using special meta-rules, and via dynamic
constraints that are also able to consider partially specified sequences of events that happened, or that are expected to happen or not to happen. 
The experiments, performed in the DALI language, have shown that the proposed approach is computationally affordable. The complexity of check has been evaluated and discussed. We have argued and shown by means of examples that the proposed toolkit is applicable to the field of Machine Ethics, in particular to check and enforce ethical behaviour in intelligent agents.

There are relationships between our approach and event-calculus formulations, e.g., the ones presented in
\cite{GanasciaT14} where however the temporal aspects and the correction of violations are not present. Approaches based on abductive logic programming such as
SCIFF (cf. \cite{sciff2011} and the references therein) allow one to
model dynamically upcoming events, and to specify positive and negative expectations, together with the concepts of
fulfilment and violation of expectations.
Reactive Event Calculus (REC) stems from SCIFF \cite{sciff-iclp2008} and adds more flexibility by
reacting to new events by extending and revising previously computed results.
These approaches have been devised for either static or
dynamic checking, the latter however performed by a third party and not fully integrated within the agent's operation like in the present proposal. Event sequences, the concepts
of violated and broken expressions, complex reaction patterns,
and independence of the underlying logic are other distinguished features of our approach, never proposed before. 

A well-established line of work concerning the use of temporal logic in order to define
run-time monitors is discussed in \cite{BarringerRH10} and the references therein.
However, this work is not related to agents, and does not concern self-checking:
in fact, they propose a rule-based temporal language for defining \as monitors'' which examine either on-line or off-line some kind of \as observable trace'' generated by the program under check. There is no notion of recovery in case malfunctioning is detected. 

Deontic logic has been used for building well-behaved agents (c.f., e.g., \cite{BringsjordAB06}). However, expressive deontic logics are undecidable\footnote{The author wishes to acknowledge former Ph.D. student Abeer Dyoub for the thorough investigation of the applications of deontic logic to build ethical agents during the development of her Thesis \cite{DyoubThesis2019}.}. Therefore, although our approach cannot compete in expressiveness with deontic logics, it can be usefully exploited in practical applications. 

The proposed approach has also been experimented in the context of
energy management in smart buildings \cite{CaianielloCGFG13}. In this application domain, forms of intelligent control are needed which are
dynamic by nature, and must fulfil real-time requirements: in fact, each building has its own dynamic thermo-physical behaviour and
is immersed in a dynamic environment where weather events change its energy `footprint' in function of time. In addition, there are users' needs and preferences concerning the most suitable and comfortable temperature in each room of the building. The outcome of the experiments is encouraging, in the sense that adopting agents equipped with the proposed features allows for not only general
but also local (room-by-room or area-by-area) control of energy saving according to user comfort requirements and preferences.
 
Future work includes making self-checking constraints adaptable to changing conditions, and devising a useful integration and synergy with declarative a priori verification techniques. As suggested in \cite{Rushby08}, a very interesting line of investigation concerns automated synthesis of runtime constraints from specifications but also from test results, extracting invariants expressing correct or critical situations.

An unsolved issue in our setting is
explicit treatment of time. In fact, in Evolutionary A-ILTL expressions time is treated implicitly
by means of the sequence of states underlying the interval temporal logic. These states are related
to the subjective agent's perception of events, and on the total ordering of their time-stamps.
Investigating how to incorporate in the approach a more general representation of \as real'' time, deadlines, etc.
is another subject of future work.

Finally, we intend to attempt a synergy between this approach and our recent line of work on learning ethical rules. More broadly, we would like to extend the approach to learning agents in general.

\end{sloppypar}
\end{document}